\documentclass[11pt]{article}

\usepackage{epsfig}

\newcommand{\sect}[1]{\setcounter{equation}{0}\section{#1}}

\def\be{\begin{equation}}
\def\ee{\end{equation}}
\def\ba{\begin{eqnarray}}
\def\ea{\end{eqnarray}}

\title{{\bf Capture of bulk geodesics by brane-world black holes}}
\author{Andrew Chamblin\thanks{email: chamblin@mit.edu} \\ 
Center for Theoretical Physics, \\
MIT, Bldg. 6-304, \\ 
77 Mass. Ave. \\
Cambridge, MA 02139, U.S.A.\\ \\ }

\begin{document}

\maketitle

\begin{abstract}

In this note we study bulk timelike geodesics in the presence of a brane-world
black hole, where the brane-world is a two-brane moving in a $3+1$-dimensional
asymptotically $adS_4$ spacetime.  We show that for a certain range of the
parameters measuring the black hole mass and bulk cosmological constant, 
there exist stable timelike geodesics which orbit the black hole and remain 
bound close to the two-brane.  

\end{abstract}

\sect{Introduction: Brane-world black holes}

As is well known, phenomenological considerations have led people
to consider the possibility that there may exist extra dimensions of space 
which are quite large.  Thus, in this framework the 
universe would be a three-dimensional brane moving in some higher dimensional
spacetime.  This sort of picture is naturally realized in the
Ho\v{r}ava-Witten corner of the M-theory moduli space, and it
is also the basic assumption underlying the second model of Randall and Sundrum
\cite{rs2},
where the universe itself is a thin ``distributional" static flat
domain wall or three-brane
separating two regions of five-dimensional anti-de-Sitter spacetime.

Of course, we need to check that this idea that we live on a brane
is consistent with what we already know.  For instance,
if matter trapped on a brane undergoes gravitational collapse then a
black hole will form. Such a black hole will have a horizon that
extends into the dimensions transverse to the brane: it will be a
higher dimensional object.  Consistency demands that this higher
dimensional object reproduce the usual
astrophysical properties of black holes and stars.

Motivated by this reasoning, various authors have studied the problem
of gravitational collapse on the brane.  In particular, in \cite{bh}
the authors proposed the existence of a `black cigar' solution, and
conjectured that this black cigar
solution is the unique stable vacuum solution in five dimensions
which describes the endpoint of gravitational collapse in a 
Randall-Sundrum brane-world.  However, though they were able to 
discuss the gross features of this solution, they were not able
to write the metric down explicitly.  Since this work, other papers
have appeared generalizing the analysis to include the effects of charge,
and also the effects of more than one large extra dimension \cite{general}.

At present, about the only thing we know about 
the black cigar metric is that it is a generalization to higher dimensions
of a solution in four dimensions known as the
{\it adS-C-metric}.  This is because the adS-C-metric naturally
describes the pair creation of black holes by the breaking of a cosmic string
in $3+1$ dimensions, and the black cigar metric should describe the 
fragmentation (due to the Gregory-Laflamme instability \cite{ruth}) of 
a {\it black} string solution in $4+1$ dimensions.\footnote{The key point
here is that in $3+1$ dimensions a self-gravitating Nambu string 
generates a conical deficit in spacetime, whereas in $4+1$ dimensions
it actually generates an extended one-dimensional horizon, or
`black string'.}  Motivated by this observation, the authors of
\cite{ehm} studied the adS-C-metric, with a hope of learning something
about the geometrodynamics of brane-worlds in one less dimension.
We now turn to their construction.

\sect{The construction of Emparan, Horowitz and Myers}

In \cite{ehm} the authors studied a lower-dimensional version
of the Randall-Sundrum model, consisting of a two-brane moving
in a $3+1$-dimensional asymptotically $adS_4$ background.  They
explicitly showed how the adS-C-metric describes a black hole
bound to the two-brane, and therefore were able to conclude
that adS-C describes the final state of gravitational collapse
on the brane-world.  At first glance, this seems to go against 
the conventional wisdom that gravitational collapse in $2+1$ dimensions
(without a cosmological constant) will always produce a conical
deficit, and not a horizon.  However, this reasoning is flawed because
matter which is trapped on the brane will still generate 
curvature in the full $3+1$ dimensional spacetime.  Consequently,
collapse on the brane will produce a black hole in $3+1$ dimensions,
although at large distances along the brane $2+1$-gravity is reproduced
in the sense that there is still a deficit angle in asymptotia.
Following \cite{ehm}, we will only consider the following
special case of the adS-C-metric:
\begin{equation}
ds^2={\l^2\over (x-y)^2} \left[-(y^2+2\frac{m}{l} y^3) dt^2 +
{dy^2\over (y^2+2\frac{m}{l} y^3)} + {dx^2\over G(x)} +G(x) d\varphi^2 \right]
\label{cads}
\end{equation}
where $G$ is the cubic polynomial
\[
G(x)=1- x^2 -2 \frac{m}{l} x^3
\]
This metric is a solution of the bulk Einstein equations with a 
negative cosmological constant
\[
\Lambda =  - 6/l^2
\]
The global structure of this solution was more than adequately 
discussed in \cite{ehm}; here, we simply recall the salient points.

First, the acceleration parameter $A$, which measures the rate
at which the black hole has to accelerate in order to remain bound to 
the brane-world, has been fine tuned relative to the bulk cosmological
constant:
\begin{equation}
A = 1/l
\end{equation}
Of course, this is just a reflection of the fact that we are working
in the original second Randall-Sundrum scenario, where the brane tension
is fine tuned relative to the bulk cosmological constant so that the
worldvolume of the brane is Minkowski.  Any black hole which comoves
with a Minkowski brane in adS has to accelerate at the rate (2.2).
On the other hand, if the brane is de Sitter the acceleration will
be even greater.  It would be interesting to study this more general
case where the black hole is bound on a de Sitter brane-world.

Next, it is worth pointing out that `infinity' corresponds to the
set of points specified by $x = y$.  This is because these points are
infinitely far away from any other points.  Not much is known about
this infinity, except that it is still a timelike surface where presumably
a holographic theory may reside.  The structure of this set is not
relevant for the construction of \cite{ehm}, since it is precisely this
part of the spacetime that we `throw away' when we cut-and-paste to obtain
the Randall-Sundrum domain wall.

The coordinates used in (2.1) are probably not familiar to most people.
Basically, the angle $\varphi$ rotates around the axis of symmetry of the
black hole (i.e., it rotates around the cosmic string which is pulling
the black hole towards infinity).  The coordinate $x$ is like the
cosine of the other angular variable, and the coordinate $- y$ is a
radial variable.  In this paper, we are going to be interested in 
geodesics which `{\it co-accelerate}' with the black hole \cite{pravda} as it is 
pulled towards infinity.  That is, we want to see if there can exist
geodesics which stay near the equatorial plane of the black hole
($x=0$, where the brane-world intersects the black hole), so that $x$ = constant,
and which simultaneously orbit the black hole at some constant
radius $y$ = constant.

As usual, there is a horizon wherever the metric component $g_{tt}$ has
a root.  The root at $y = 0$ corresponds to the acceleration horizon for
the black hole, or equivalently, to the bulk adS Cauchy horizon.  The 
horizon at $y = -l/2m$ is the black hole horizon.  There is a curvature
singularity at $y = - \infty$.

The parameter $m$ measures the mass of the black hole, and as discussed
in \cite{ehm} once we specify $m$ and $l$, the size of the deficit angle
${\Delta}{\varphi}$ is fixed.  This deficit angle measures the tension
of the `cosmic string' which is accelerating the black hole towards the
boundary at infinity.  In the construction of Emparan et al, we throw
away the part of the spacetime containing the conical deficit and we will
have nothing more to say about this portion of the spacetime.

Finally, we should mention that we will assume the metric is Lorentzian
signature $(- + + +)$ throughout, and consequently we must have
$G(x) ~{\geq}~ 0$.  Also, we will require that $G$ possesses three
distinct real roots, so that $0 < \frac{m}{l} < \frac{1}{3\sqrt{3}}$ as in \cite{ehm}.

\sect{Analysis of the geodesics}

In \cite{pravda}, Pravda and Pravdova studied geodesic motion in the C-metric,
which is a solution of the vacuum Einstein equations with vanishing
cosmological constant.  The following analysis is a straightforward
generalization of their work to the adS-C-metric.

The lagrangian associated with the metric (2.1) is given as
\begin{equation}
L =\frac{l^2}{(x - y)^2}\left(
           +\frac{1}{G}(\frac{{\rm d}x}{{\rm d}\tau})^2
           +\frac{1}{F}(\frac{{\rm d}y}{{\rm d}\tau})^2
           +G(\frac{{\rm d}\varphi}{{\rm d}\tau})^2
           -F(\frac{{\rm d}t}{{\rm d}\tau})^2\right)
\end{equation}
where $F(y) = y^2 + 2\frac{m}{l} y^3$, $\tau$ is the proper time experienced by a free falling
timelike observer (and the affine parameter for null geodesics), 
$L = -1$ for timelike geodesics and $L = 0$ for null geodesics.  

The line element (2.1) is boost and rotation symmetric, that is to say,
${\partial}_t$ and ${\partial}_{\varphi}$ are Killing vectors.  As described
in \cite{pravda}, these Killing vectors define constants of the motion
$J$ and $E$ such that for any geodesic 
$\lambda(\tau) = (t(\tau), y(\tau), x(\tau), \varphi(\tau))$
we have the relations
\begin{equation}
\frac{{\rm d}{\varphi}(\tau)}{\rm d \tau}  =
     \frac{J}{l^2} \frac{[ x(\tau ) - y(\tau )]^2}
              {G(x(\tau ))}
\end{equation}
and
\begin{equation}
\frac{{\rm d} t(\tau)}{\rm d \tau} = 
     \frac{E}{l^2} \frac{[ x(\tau ) - y(\tau )]^2}
               {F(y(\tau ))}
\end{equation}

\subsection{Timelike geodesics}

If we substitute equations (3.2) and (3.3) into (3.1), and let $L = -1$, we obtain
the equation
\begin{equation}
\frac{F l^4}{E^2(x - y)^4}\left[ 
        \frac{1}{G}( \frac{{\rm d}x}{{\rm d}\tau})^2
       +\frac{1}{F}( \frac{{\rm d}y}{{\rm d}\tau})^2\right]
    = \frac{E^2 - V^2}{E^2}
\end{equation}
where $V$ is an effective potential for the motion given as
\begin{equation}
V = F^{1/2} \left( \frac{l^2}{(x - y)^2} + \frac{J^2}{G} \right)^{1/2}
\end{equation}
Clearly, (3.4) is precisely what we want, given that we are interested
in the stability of timelike geodesics that orbit at $x =$ constant,
$y =$ constant.  In particular, we want to know if there can 
exist geodesics with $x =$ constant and $y =$ constant in a region where
the potential $V$ has a local minimum.  As in \cite{pravda}, 
we find that such a local minimum for $V$ can
exist, provided the mass $m$ and length scale $l$ satisfy the 
necessary inequality:
\begin{equation}
\frac{m}{l} < C
\end{equation}
where $C$ is some constant which has to be determined numerically.
The precise value of this constant is difficult to estimate, although
our results imply that $C ~{\sim}~ {\cal O}(10^{-3})$.

In other words, if a black hole does {\it not} satisfy (3.6) then
there is definitely no local minimum for $V$ and a slight perturbation
will cause the particle to either fall away from the brane into the bulk
or into the black hole.  Conversely, if (3.6) is satisfied then it is
possible to find geodesics with $x$ and $y$ constant.  Generically,
the stable orbits occur for a range of $y$ outside of the black hole
horizon: $-\frac{l}{2m} < y_{min} < y < y_{max} < 0$.

We are not able to find any geodesic which
orbits {\it exactly} at $x = 0$, although we are able to find such
orbits with $x$ arbitrarily small.  This is perhaps unsurprising,
given the symmetry of the setup and the fact that we are trying to
balance bulk acceleration vs. black hole acceleration.  It would be interesting to know
if it is ever possible to satisfy $x = 0$ for a stable orbit, perhaps
by including the effects of charge or rotation.

\subsection{Null geodesics}

As in \cite{pravda}, it is straightforward
to show that there exists a {\it null} geodesic ($L = 0$) which
stays bound to the brane (at $x(\tau) = 0$), and which orbits
the black hole at a radius
\[
y(\tau) = - \frac{l}{3m}
\]
Thus, we see explicitly how the adS length scale $l$ will affect
the radius at which there may exist a null geodesic orbit.
As in \cite{pravda}, however, this orbit is unstable and the
slightest perturbation will cause the photon to either fall into
the black hole or off of the brane-world and into the bulk.

\sect{Conclusion: Do some black holes have `halos'?}

We have shown that a brane-world black hole can be used to 
`trap' bulk degrees of freedom, in the sense that there can exist
stable bulk geodesic orbits which remain bound 
arbitrarily close to the brane
as they orbit the black hole.  Of course, we have only
really seen that this is true for a two-brane moving in 
an $adS_4$ background, but for the sake of argument let us
suppose that a similar result holds for the full black cigar
metric.  

It would then follow that for any black hole which satisfies 
a bound of the form (3.6)
there exists a sort of `halo' surrounding the black hole where particles
can orbit the black hole in such a way that the total acceleration
5-vector is nearly vanishing.  Crudely, what is happening is 
that the acceleration due to the black hole is compensating for the acceleration
due to the adS bulk, in such a way that in the halo region
particles can orbit in inertial frames.  Naively, one might 
conclude that this would imply
that the clock of a brane-world observer would run {\it faster} inside
of the halo than it would outside of the halo \cite{sean}.  This is because an 
observer orbiting far outside of the halo region will have a non-vanishing
acceleration $A_{outside}$ (the acceleration
required to stay on a horospherical brane in adS), but an observer orbiting
inside of the halo can arrange for her total acceleration to be arbitrarily small:
\[
A_{halo} ~{\sim}~ 0
\]
However, this reasoning is incorrect because it presupposes that acceleration
makes sense on a spacetime that has a metric discontinuity.  More precisely,
the assumption that the brane-world is infinitely thin is merely an
idealization, which implies that the acceleration experienced by observers
who comove with the brane-world must jump discontinuously as one moves
through the brane.  To avoid this discontinous jump, 
the spacetime must somehow be `smoothed out'
in the region of the domain wall.  Such a smoothing will imply that there
is a region in the center of the domain wall where the
force experienced by comoving observers is zero.  Only by moving a certain
critical distance from the brane will observers begin to feel the
acceleration which drags them into the bulk of adS.  In this way we
can avoid the sort of paradoxical  behaviour discussed above.

On a more speculative note, it is tempting to draw an analogy 
between the halo where bulk degrees of freedom might reside and the 
galactic halo of `dark matter'.  Unfortunately, such a comparison seems
unlikely given that the bound (3.6) is an {\it upper} bound on the mass
of the black hole.

On the other hand, it certainly does make sense
to think of these trapped bulk degrees of freedom in terms of the effective
theory on the brane-world, as discussed in \cite{gk}.
From this perspective, it is clear that a bulk particle which is
trapped in the halo region will have a
holographic image on the brane with some spread - a kind of cloud.
What our results imply is that the bulk particles orbiting in the 
halo will generate an orbiting cloud of dark matter, which can interact
with brane-world matter through the exchange of gravitons
and SYM gauge bosons.
One might think that in order to pursue this line of reasoning we need
the full black cigar metric; however, it is likely that some 
linearized analysis, such as that presented in \cite{gkr},
will do the trick.  Research on this and related issues is
currently underway.

{\noindent \bf Acknowledgements}\\

I thank R. Emparan, S. Giddings, A. Guth, E. Katz and 
N. Lambert for useful discussions, and the faculty and
staff of the High Energy Theory Group at the University of
Pennsylvania for hospitality while this work was completed.  This
work was supported in part 
by funds provided by the U.S. Department of Energy (D.O.E.) under
cooperative research agreement DE-FC02-94ER40818.

\end{document}